\documentclass{elsart3p}
\usepackage{graphics}
\usepackage{graphicx}
\usepackage{amssymb}

\newcommand{\kvec}{{\bf k}}
\newcommand{\qvec}{{\bf q}}

\newcommand{\Kvec}{{\bf K}}
\newcommand{\Qvec}{{\bf Q}}

\begin{document}

\begin{frontmatter}

\title{Kinks and d-waves from phonons: The intermediate coupling story}

\author[label1]{J.P.Hague}
\ead{J.P.Hague@open.ac.uk}
\corauth{Tel: +44-1509-228208.  Fax: +44-1509-223986}
\address[label1]{Department of Physics and Astronomy, The Open University, Milton Keynes, MK7 6AA, UK}
\thanks[epsrc]{I acknowledge support by EPSRC grant EP/C518365/1}

\begin{abstract}
I present results from an approach that extends the Eliashberg theory
by systematic expansion in the vertex function; an essential extension
at large phonon frequencies, even for weak coupling. In order to deal
with computationally expensive double sums over momenta, a dynamical
cluster approximation (DCA) approach is used to incorporate momentum
dependence into the Eliashberg equations. First, I consider the
effects of introducing partial momentum dependence on the standard
Eliashberg theory using a quasi-local approximation; which I use to
demonstrate that it is essential to include corrections beyond the
standard theory when investigating $d$-wave states. Using the extended
theory with vertex corrections, I compute electron and phonon spectral
functions. A kink in the electronic dispersion is found in the normal
state along the major symmetry directions, similar to that found in
photo-emission from cuprates. The phonon spectral function shows that
for weak coupling $W\lambda < \omega_0$, the dispersion for phonons
has weak momentum dependence, with consequences for the theory of
optical phonon mediated d-wave superconductivity, which is shown to be
2nd order in $\lambda$. In particular, examination of the order
parameter vs. filling shows that vertex corrections lead to $d$-wave
superconductivity mediated via simple optical phonons. I map out the
order parameters in detail, showing that there is significant induced
anisotropy in the superconducting pairing in quasi-2D systems. {\bf [PUBLISHED AS: JOURNAL OF PHYSICS AND CHEMISTRY OF SOLIDS, VOL. 69, 2982-2985 (2008)]}
\end{abstract}

\end{frontmatter}

\section{Introduction}

Angle-resolved photo-emission spectroscopy (ARPES) directly probes the
dispersion of electrons; identifying a kink associated with the
active optical phonon in the cuprates \cite{lanzara2001a}. Neutron scattering has shown an anomalous change in the phonon
spectrum at the superconducting transition, indicating an interesting role for phonons in the cuprates \cite{chung2003a}. Estimates of the magnitude
of the electron-phonon coupling, and the energy of the phonon mode put
the problem outside the limited region of applicability for BCS
theory, so schemes to cope with a wider range of parameters need to be
developed.  Moreover, since there a node in the superconducting gap
consistent with d-wave symmetry \cite{tsuei1994a}, any theory
implicating phonons as the mechanism must be able to explain the
unconventional order.

Electron-phonon (e-ph) interactions can be described by the following generic model,
\begin{eqnarray}
H &=& \sum_{\kvec\sigma}\epsilon_{\kvec}c^{\dagger}_{\kvec,\sigma} c_{\kvec,\sigma} + \sum_{kq} \frac{g_{\qvec}}{\sqrt{\omega_{\qvec}}} c^{\dagger}_{\kvec-\qvec,\sigma}c_{\kvec,\sigma} (b^{\dagger}_{\qvec}
 + b_{\qvec})\nonumber\\ 
&& \hspace{10mm} + \sum_{\qvec} \omega_{\qvec} (b^{\dagger}_{\qvec}b_{\qvec} + 1/2)
\end{eqnarray}
Here, $\hbar = 1$ and $\epsilon_{\kvec} = -2t(\cos(a k_x)+\cos(a k_y)) -
2t_{\perp}\cos(c k_z)$ with $t=0.25$ and $t_{\perp}=0.01$
(representing a quasi-2D system and taming the van-Hove
singularities). The Holstein model is a special case where $g_{\qvec}
= g_0$ and $\omega_{\qvec} = \omega_0$, for which $\lambda =
g_0^2/(\omega_0^2 W)$ is the dimensionless e-ph coupling
($W$ is the half-band width). The momentum-independent phonon
dispersion approximates an optical phonon, and the momentum-independent e-ph coupling corresponds to a completely local
interaction. A Coulomb pseudo-potential, $H_\mu = -
W\mu_c\sum_{\kvec\qvec\sigma} \langle
c^{\dagger}_{\qvec\sigma}c^{\dagger}_{-\qvec\bar{\sigma}} \rangle
c_{\kvec\sigma} c_{-\kvec\bar{\sigma}} + h.c.$, may be added
(i.e. Hubbard model at the Hartree--Fock level).

\begin{figure}[ht]
\begin{center}
\includegraphics[width=0.46\textwidth]{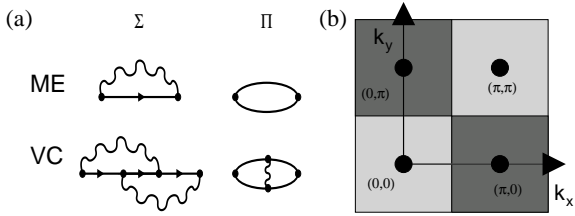}
\end{center}
\caption{\label{fig:overview}(a) Overview of the diagrammatic
  representations of the self-energy used in this work. $\Sigma$ is
  the electron, and $\Pi$ the phonon self energy. The row labeled ME
  contains the vertex neglected and VC the corrected diagrams. (b)
  Schematic of the DCA (quasi-local) formalism for cluster size
  $N_C=4$. Within the squares $\Sigma(\kvec)\mapsto\Sigma(\Kvec)$
  where $\Kvec$ are represented by solid circles. Equivalently,
  $\Delta(\kvec)\mapsto\Delta(\Kvec)$. A partial density of states
  (DOS) is associated with each sub-zone. DCA enables computations in
  the thermodynamic limit, via an equivalent cluster impurity
  problem.}
\end{figure}

\begin{figure*}[ht]
\begin{center}
\includegraphics[height=0.325\textwidth,angle=270]{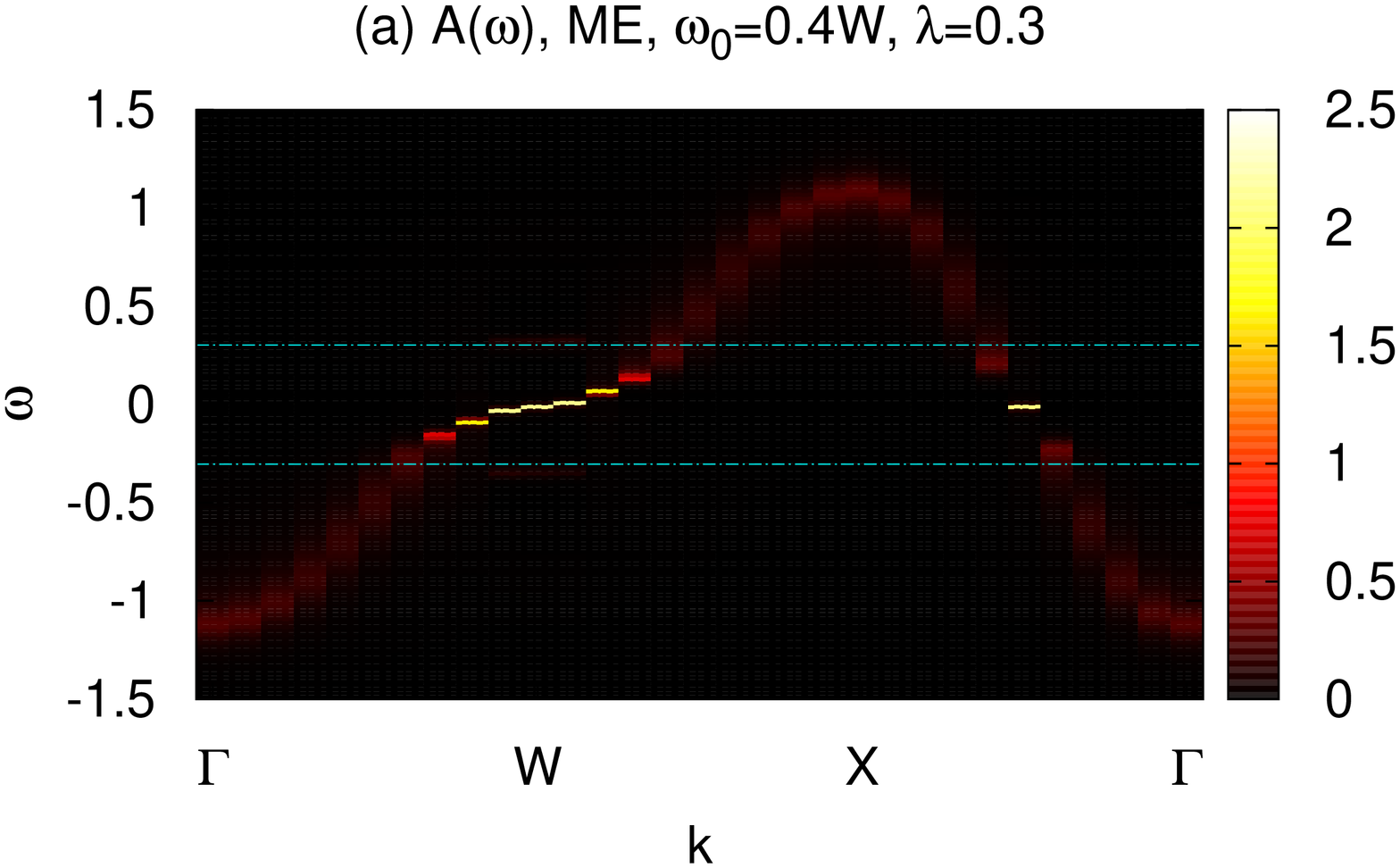}
\includegraphics[height=0.325\textwidth,angle=270]{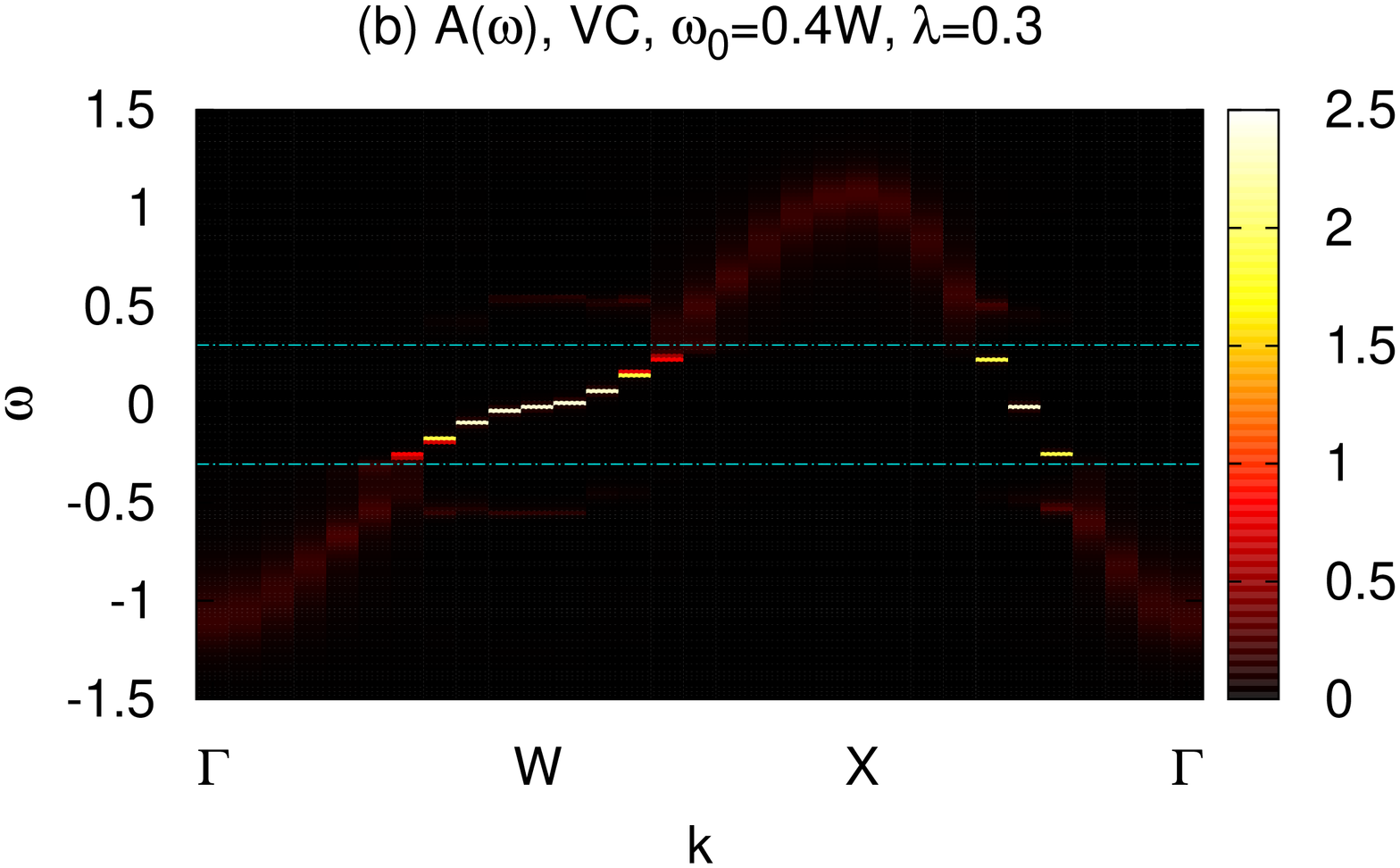}
\end{center}
\caption{\label{fig:one} (Colour online) Electron spectral functions
  comparing VC and ME approximations when cluster size $N_{C}=64$. The
  effects of vertex corrections are most pronounced for
  $\omega_0>W\lambda$, as expected from analysis of the parameter
  space \cite{haguenda}. Horizontal dashed lines indicate the
  renormalised phonon frequency at the $(\pi,0)$ zone-edge as computed
  within the self-consistent theory (i.e. the one which would be
  measured in a neutron experiment). In particular, only the vertex
  corrected theory shows a kink at the energy scale relating to the
  phonon frequency {\bf measured} at the $(\pi,0)$ point
  \cite{hague2003a} as reported in Ref. \cite{lanzara2001a}.}
\end{figure*}

Some treatments of the ARPES kink used a standard Eliashberg theory
with anisotropic coupling and an assumed $d$-wave order parameter for
fitting purposes \cite{devereaux2004a,cuk2004a}. The extended Eliashberg
approach that I have developed differs significantly from previous
work in the sense that it is fully self-consistent, with no prior
assumptions about the form of the order parameter (i.e. any $d$-wave
order arises directly from the self-consistency of the theory)
\cite{hague2005d,hague2006b}.

\section{Eliashberg theory and momentum: Failure for optical-phonon mediated $d$-wave states}

In this section, I discuss how quasi-local Eliashberg equations can be
constructed, and demonstrate how the lowest-order Fock diagram does
not contribute to the gap equations. The calculation is intended to be demonstrative, and shows that standard
Eliashberg equations fail for an optical phonon mediated $d$-wave
state.

The Eliashberg equations are computed from the
Fock contribution to the self-energy:
\begin{eqnarray}
\hat{\Sigma}(\kvec,i\omega_n) & = & -T\sum_{\omega_{n'}}\int \frac{d\qvec}{(2\pi)^3}g^2_{\qvec}\hat{\tau}_3\hat{G}(\kvec-\qvec,\omega_{n'})\nonumber\\
& & \hspace{10mm}\times\hat{\tau}_3 K(\qvec,i\omega_n-i\omega_{n'})
\label{eqn:se}
\end{eqnarray}
and equate $\hat{\Sigma}$ with the approximate form,
\begin{equation}
\hat{\Sigma}\approx (1-Z_{\kvec})i\omega_{n}\hat{\tau}_0 + \Delta_{\kvec}(i\omega_n)\hat{\tau}_1 + \chi\hat{\tau}_3
\end{equation}
Normally the Eliashberg equations are computed using a local
(momentum-independent) approximation. However, to examine $d$-wave
states, the local approximation is not sufficient. Substituting the
Green function
$\hat{G}(\kvec,i\omega_n)^{-1}=i\omega_n\hat{\tau}_0-\zeta_{\kvec}\hat{\tau}_3-\hat{\Sigma}$
in eq. \ref{eqn:se} ($\zeta_{\kvec}=\epsilon_{\kvec}-\mu$, $\hat{\tau}$ are Pauli matrices and $\mu$ is
the chemical potential), and treating the gap function as constant in
sub-zones of $\kvec$-space in the manner of DCA, one obtains a
quasi-local approximation for the Eliashberg equations,
\begin{equation}
\Delta_{n,\Kvec} = \lambda T \sum_{n'\Qvec}\int d\zeta \frac{\bar{K}(n,n'){\mathcal{D}}_{\Kvec-\Qvec}(\zeta)\Delta_{n',\Kvec-\Qvec}}{Z^2_{\Kvec-\Qvec}\omega_{n'}^2 + \zeta^2+\Delta^2_{n',\Kvec-\Qvec}}
\label{eqn:qlapprox}
\end{equation}
When gap functions have $d$-wave symmetry, $\Delta_{\pi,0} =
-\Delta_{0,\pi}$ and $\Delta_{0,0} = \Delta_{\pi,\pi} = 0$, and the
partial DOS ${\mathcal{D}}(0,\pi) = {\mathcal{D}}(\pi,0)$ has the symmetry of the lattice, it is
immediately clear that the $\mathcal{O}(\lambda)$ contribution to the
quasi-local Eliashberg equations vanishes, and 2nd order terms must be
considered. When there is some weak momentum dependence in the e-ph
coupling and phonon frequency, this cancellation will not be exact,
but the contribution from 2nd-order terms can still be expected to be
the largest in the perturbation series. Thus a fully extended set of
Eliashberg equations including vertex corrections needs to be
constructed. Currently, it is not possible to construct a similar approximation to eq. \ref{eqn:qlapprox} for the vertex corrected theory. However a numerical
approach using the full DCA equations is available to enlighten some
aspects of the e-ph problem.

The rest of this article concerns results from an extended Eliashberg
theory with momentum dependence and vertex corrections determined
numerically using DCA. No form for the order parameter is assumed in
advance and full forms for the Green functions and self-energies are
used. Full details of the translation of the diagrams in Fig. \ref{fig:overview}(a) may be found in
ref. \cite{hague2005d}.

\section{Normal state spectroscopy}

Figure \ref{fig:one} shows image plots of the electron spectral
functions \cite{hague2003a}. Vertex corrections are most pronounced for
$\omega_0>W\lambda$, as expected from analysis of the parameter space
\cite{haguenda}. In particular, the vertex corrected theory shows
a kink at the energy scale relating to the phonon frequency {\bf
measured} at the $(\pi,0)$ point (i.e. renormalised zone edge phonon
frequency) in agreement with the ARPES measurements in
Ref. \cite{lanzara2001a}.

\begin{figure*}[ht]
\begin{center}
\includegraphics[width=0.65\textwidth, height=6cm]{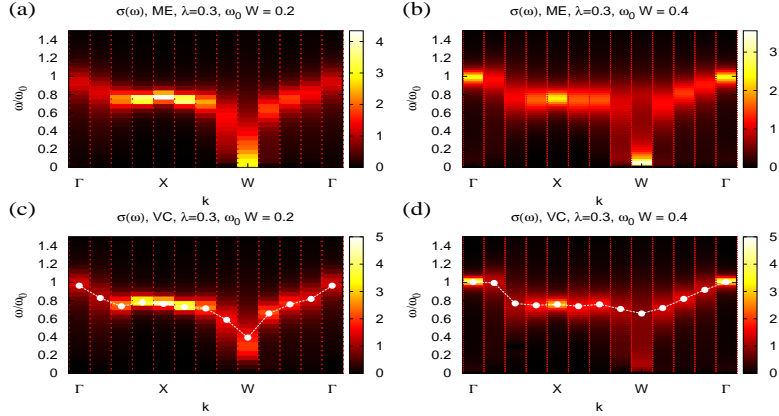}
\end{center}
\caption{\label{fig:two} (Colour online) Phonon spectral
functions. The phonon renormalisation is moderate for $\omega_0
= 0.4W$, $N_{C}=64$. White points show the values computed from $\omega^2 - \omega_0^2(1-{\rm Re}[\Pi(\omega)])=0$ for clarification. (a) Modes at the W point in
the Brillouin zone are almost completely softened in the theory
without vertex corrections (a split mode is indicated in panel (b)). Vertex
corrections act against that softening and reinforce the theory (c,d). There is an indication of reduction of spectral weight half way between the high symmetry points when $\omega_0>W\lambda$, perhaps with similar origin to that in the neutron results \cite{chung2003a}}
\end{figure*}

Phonon spectral functions are shown in Fig. \ref{fig:two}. The phonon
renormalisation is extremely small for $\omega_0 = 0.4W$. In contrast,
modes at the $(\pi,\pi)$ point in the Brillouin zone are strongly
softened for $\omega_0 = 0.2W$ when vertex corrections are not taken
into account. The second order terms act against the softening and
reinforce the theory. A small reduction in spectral weight is seen mid
way along the $\Gamma$-X and $\Gamma$-W lines when
$\omega_0>W\lambda$. This might be a precursor to the strong softening
seen in Ref. \cite{chung2003a} on passing through the transition
temperature, but further analysis of the superconducting state is
necessary to be sure.

\section{Evolution of the order parameter}

\begin{figure*}[ht]
\begin{center}
\includegraphics[width=0.8\textwidth]{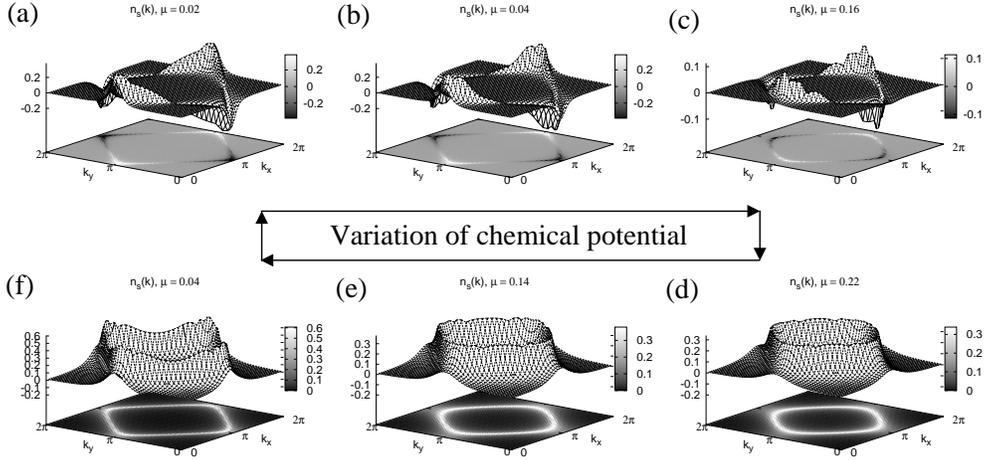}
\end{center}
\caption{\label{fig:three} Anomalous density as $\mu$ is varied first upwards from half-filling and then back from large
filling showing two types of solutions to the self-consistent
equations. $\mu_C = 1.2$, $\lambda = 0.6$, $N_C=4$ and $\omega_0/W = 0.4$. (a)
Around half-filling ($n=1$), there is only one solution with $d$-wave
order. This persists to $n=1.175$ where the order parameter
spontaneously changes to $s$-wave order (b,c). The reduction in
$d$-wave magnitude is reminiscent of a quantum critical transition.
When decreasing $\mu$ from dense filling, the order parameter is a
conventional $s$-wave (d), which persists below $n=1.175$, i.e. there
are 2 solutions (e). Below $n=1.05$ there is only one solution. Close
to $n=1.05$, the $s$-wave solution has an extended-$s$ character, as
can be seen in panel (f).}
\end{figure*}

$d$-wave superconductivity is one of the remarkable results from the
theory extended with vertex corrections. Figure \ref{fig:three} shows
the anomalous density as the chemical potential is varied first
upwards from half-filling and then back from large filling showing
that two solutions are stable. A small $\mu_c = 1.2$ is
applied. $\lambda = 0.6$ and $\omega_0 / W = 0.4$. Starting with a
half-filled solution and increasing, then decreasing the chemical
potential, first $d$-wave and then (on decreasing from high electron
density) $s$-wave solutions to the self-consistent equations are found
(as detailed in the caption of Fig. \ref{fig:three}). Highly
anisotropic solutions form as the Fermi-surface approaches the
van-Hove points. In the $d$-wave state, the first order terms are not
the leading order of the perturbation theory in $\lambda$. It was
argued in Ref. \cite{hague2006b} that this was in part due to the
nearly flat phonon spectrum, so that the phonon propagator
$D(\kvec,\omega)\approx D(\omega)$ and the 1st order contribution to
the anomalous self-energy $W\lambda\sum_{\qvec}
D(\qvec)F(\kvec-\qvec)$ is small ($F$ is the anomalous Green
function). The anomalous contribution to the pseudopotential term is
$\phi_C = W\mu_C \sum_\kvec F(\kvec)$. Since $F$ is modulated in the
$d$-wave state, then the contribution of the pseudopotential term is
zero. In the $s$-wave state, $F$ is not modulated, and the resulting
finite contribution destroys the $s$-wave order for large enough
$\mu_C$, thus selecting a dominant $d$-wave contribution. The phonon
spectra in Fig. \ref{fig:two} demonstrate that so long as $\omega_0 >
W\lambda$, then this condition is met, and 2nd order terms are
essential. 2nd order terms remain the leading order correction to the
anomalous self-energy, even for the larger $\lambda$ used here. It is
clear that the 2nd order terms are similar for both repulsive and
attractive models, and thus it is perhaps not surprising that both
attractive and repulsive models show a $d$-wave state.

\section{Summary and Outlook}

I have described spectroscopic results and order parameters from an
extended Eliashberg-style theory with vertex and momentum
corrections. The extended theory is essential for the investigation of
d-wave superconductivity mediated by optical phonons, since the vertex
corrections are the lowest order terms in the perturbation expansion
in electron-phonon coupling $\lambda$. Indeed, d-wave
superconductivity is immediately predicted with such an extension.

There are two remaining areas which require further investigation. The
first involves a better treatment of the Coulomb repulsion, which is
currently only treated as a simple Hartree-Fock pseudopotential. In
particular, the ability to treat large repulsion and
antiferromagnetism is essential for high-$T_C$ materials. Treatment of
momentum dependent e-ph coupling would also be of interest. However,
the overriding challenge at this stage is to convert the numerical
results into a coherent analytic theory with vertex corrections.



\end{document}